\documentclass{ws-p8-50x6-00}

\begin{document}

\title{Low x Physics at HERA}

\author{A.M.Cooper-Sarkar}

\address{Particle and Astrophysics, Keble Rd, Oxford, OX1 3RH,
 UK\\ 
E-mail: a.cooper-sarkar1@physics.ox.ac.uk}


\maketitle

\abstracts{Recent HERA data on structure functions and reduced cross-sections
are presented and their significance for our understanding of the low-$x$ 
region is dicussed}

In the course of the last year both ZEUS and H1 have presented data
(see refs.~\cite{h1967},~\cite{z9697}) on structure functions and reduced 
cross-sections from the 1996/7 runs of $e^+ p$ interactions. The kinematics
of lepton hadron scattering is described in terms of the variables $Q^2$, the
invariant mass of the exchanged vector boson, Bjorken $x$, the fraction
of the momentum of the incoming nucleon taken by the struck quark (in the 
quark-parton model), and $y$ which measures the energy transfer between the
lepton and hadron systems.
The cross-section for the process is given in terms of three structure 
functions by
{\small \begin{equation}
\frac {d^2\sigma (e^+p) } {dxdQ^2} =  \frac {2\pi\alpha^2} {Q^4 x}
\left[Y_+\,F_2(x,Q^2) - y^2 \,F_L(x,Q^2)
- Y_-\, xF_3(x,Q^2) \right],
\label{eq:NCxsec}
\end{equation}}
where $\displaystyle Y_\pm=1\pm(1-y)^2$, and we have ignored mass terms.
The new data have 
extended the measured region in the $x,Q^2$ plane to cover $10^{-6} < x < 0.65$
and $ 0.045 < Q^2 < 30000 GeV^2$. The precision of measurement is such that
systematic errors as small as  $\sim 3\%$ have been achieved for 
$2 < Q^2 < 800 GeV^2$, with much smaller statistical errors. Thus the HERA
data rival the precision of fixed target data, and there is now complete
coverage of the kinematic plane over a very broad range.
In Fig.1 we show a subsample of the HERA $F_2$ 
data in comparison 
to fixed target data, for low  $Q^2$ values which cover the interesting low $x$
region.
\begin{figure}{\small Fig.1: HERA $F_2$ data compared to fixed target data at low $Q^2$}
\epsfxsize=20pc
\epsfbox{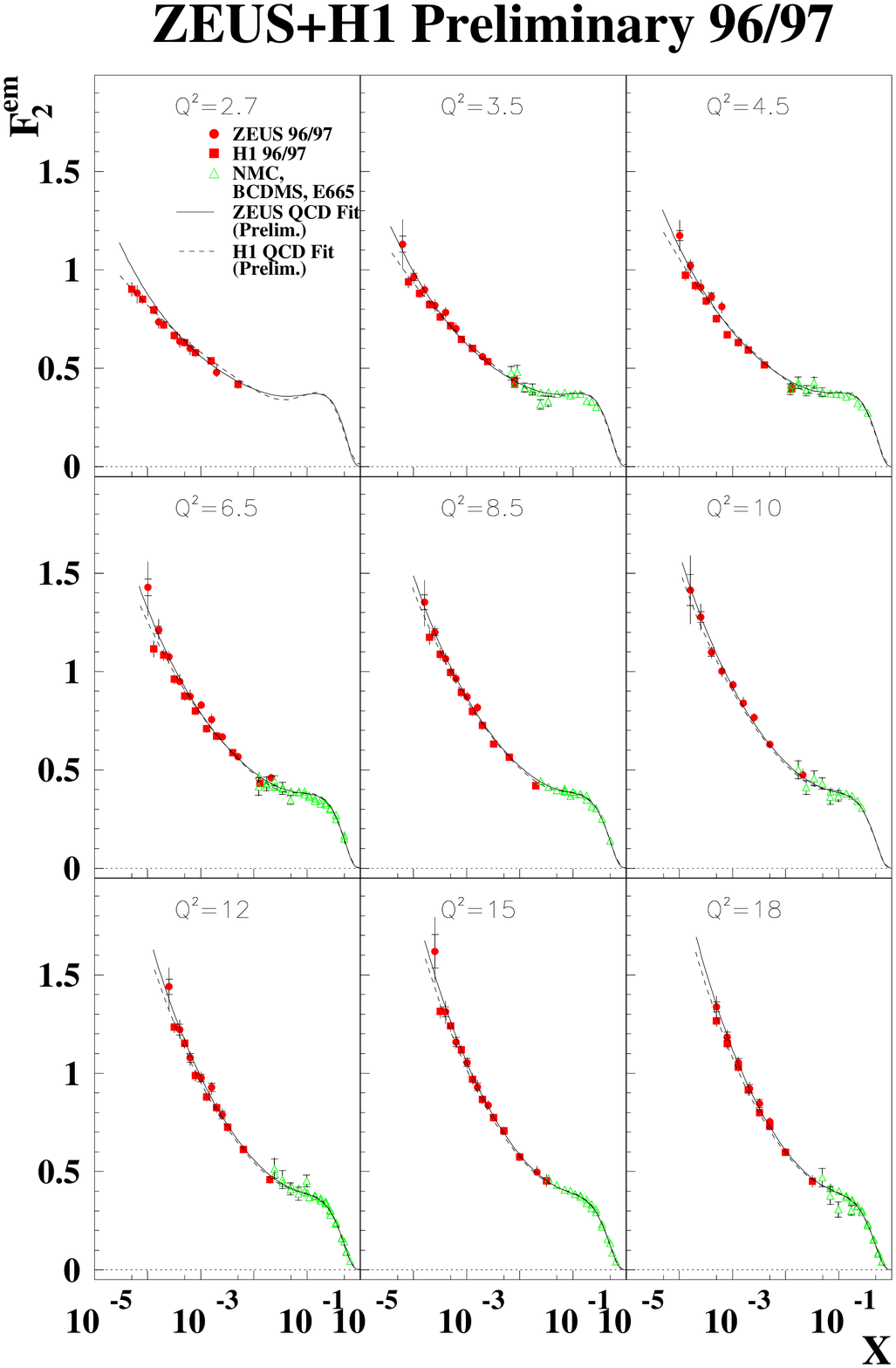}
\label{fig:f2low}
\end{figure}
This plot show the characteristic rise of $F_2$ at small $x$ which becomes more
dramatic as $Q^2$ increases. In this kinematic region, the parity
violating structure function $xF_3$ is
negligible and the structure functions $F_2, F_L$ are given purely by
$\gamma^*$ exchange. At leading order (LO) in perturbative QCD, 
$F_2$ is given by
{\small \begin{equation}
        F_2^{ep}(x,Q^2) = \Sigma_i e_i^2*(xq_i(x,Q^2) + x\bar q_i (x,Q^2)),
\end{equation}}
a sum over the (anti)-quark momentum
distributions of the proton multiplied by the corresponding 
quark charge 
squared $e_i^2$. At the same order, the spin-1/2 nature of the quarks implies 
that $F_L = 0$, thus cross-section data measure $F_2$ and 
tell us about the behaviour of the quark 
distributions, and futhermore, their $Q^2$ dependence, or scaling violation, 
is predicted by pQCD. Preliminary NLO pQCD fits to the $F_2$ data from each 
of the collaborations are shown on Fig.1.

To appreciate the significance of the QCD scaling violations 
we also show the HERA96/7 data as a function of $Q^2$ 
in fixed $x$ bins in Fig.2.
\begin{figure}{\small Fig.2: ZEUS and fixed target $F_2$ data as a function of $Q^2$ in fixed $x$ bins}
\epsfxsize=25pc
\epsfbox{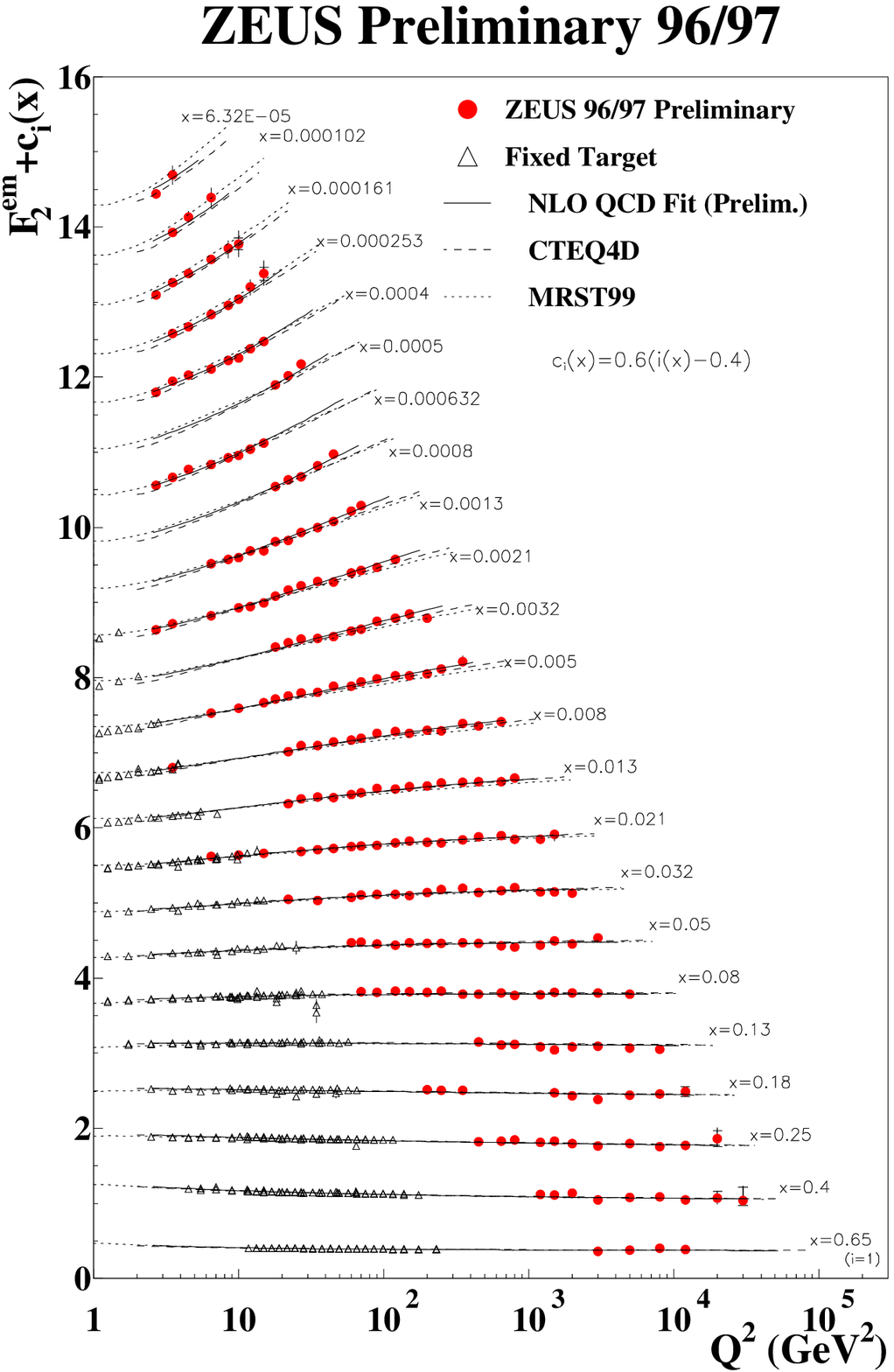}
\label{fig:scalvio}
\end{figure}
Such data has been used to extract parton distributions using
an NLOQCD fit to the DGLAP equations. For example,
{\small \begin{equation}
\frac {d q_i(x,Q^2)} {d \ln Q^2} = \frac {\alpha_s(Q^2)} {2\pi}
\int_x^1 \frac {dy}{y} \left[\sum_j
q_j(y,Q^2) P_{q_iq_j}(\frac{x}{y}) + g(y,Q^2) P_{q_ig} (\frac{x}{y}) \right]
\label{eq:DGLAPq}
\end{equation}}
describes the $Q^2$ evolution of a quark distribution in terms of parent
parton (either quark or gluon) distributions,
where the `splitting function' $P_{ij}(z)$ (predicted by QCD) 
represents the probability of the parent parton
$j$ emitting a parton $i$, with momentum fraction $z$
of that of the parent, when the scale changes from
$Q^2$ to $Q^2 + d \ln Q^2$. The QCD running coupling, $\alpha_s(Q^2)$, 
determines the rate of such processes.
Thus although the structure function $F_2$ is directly related to
quark distributions, we may also gain information on the gluon distribution
from its scaling violations. In fact at low $x$ the gluon contribution
dominates the evolution of $F_2$.

In recent years more emphasis has been
placed on estimating errors on extracted parton distributions. 
Fig.3 shows the gluon distribution extracted from a fit to 
H196/7 data, where the errors include not only experimental 
correlated systematic errors but also model errors, such as the uncertainty of 
$\alpha_s$, scale uncertainties etc. (see ref~\cite{h1967}).
\begin{figure}{\small Fig.3: H196/7 gluon illustrating 
experimental and model dependent errors}
\epsfxsize=15pc
\epsfbox{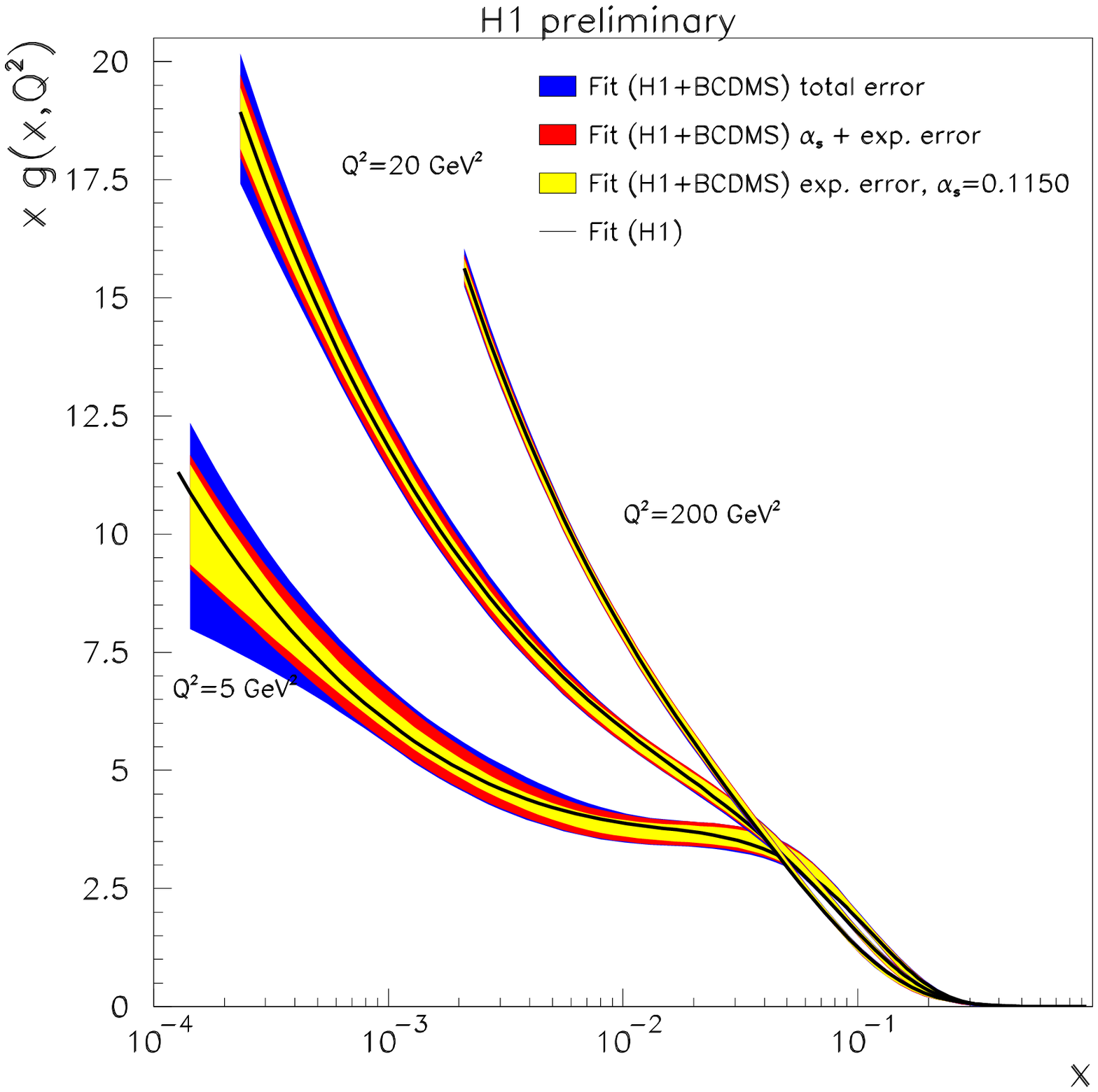}
\label{fig:h1glu}
\end{figure} 
Precision meausrements of $\alpha_s$ are also 
possible using this scaling violation
data and H1 have combined their data with that of BCDMS to obtain,
$\alpha_s=0.115 \pm 0.0017 (exp) \pm 0.0007(model) \pm 0.005(scale)$. 
It is clear that the largest uncertainties are now theoretical and that pQCD
calculations to NNLO should help to reduce this uncertainty. 

However,
when doing such fits the question arises how low in $x$ 
should one go using conventional theory? The DGLAP formalism makes
the approximation that only dominant terms in leading (and next to leading) 
$ln(Q^2)$ are resummed. However at low $x$ terms in leading 
(and next to leading) $ln(1/x)$ may well be just as important. This requires
an extension of conventional theory such as that of the BFKL resummation.
One may also question how low in $Q^2$ one should go. The DGLAP formalism
only sums diagrams of leading twist, and it is also clear that $\alpha_s$
becomes large at low $Q^2$ such that perturbative calculations cannot be used,
see ref.~\cite{amcs} and references therein 
for a full discussion of these matters.
When DGLAP fits to $F_2$ data are used to extract gluon 
distributions at  $Q^2 \leq 2 GeV^2$ one finds the 
surprising result that
the gluon becomes valence-like in shape, falling rather than rising at 
$x \leq 10^{-3}$ (see ref~\cite{Zphenom94}). 
This effect is accentuated when account is taken of NNLO 
terms~\cite{mrstnnlo}, 
when the gluon distribution may even become negative, see
Fig.4.
\begin{figure}
\epsfxsize=20pc
\epsfbox{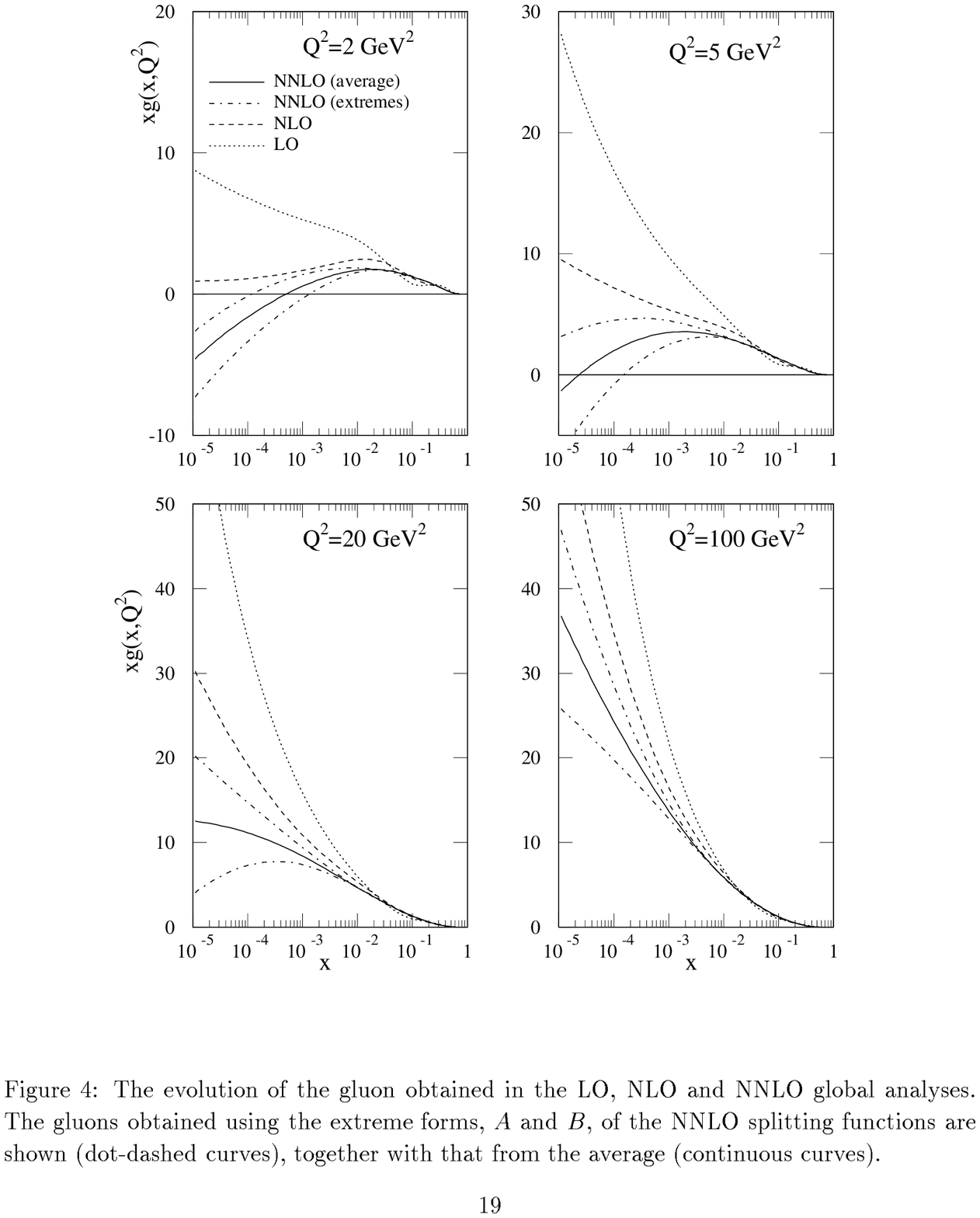}
\label{fig:mrstglu}
\end{figure} 
Such a prediction is not in itself a problem, since the gluon is not a 
physical observable, but there would be a problem if the corresponding
longitudinal structure function $F_L$ were to be negative. 
At NLO (and higher orders) QCD predicts that the longitudinal structure
function $F_L$ is no longer zero. It is a convolution of QCD coefficient
functions with $F_2$ and the gluon distribution such that 
at small $x$ ($x \leq 10^{-3}$) the dominant contribution comes from the
glue. The NNLO prediction for $F_L$ is not negative 
but it is still a rather peculiar shape, see Fig.5, 
where the DGLAP predictions for LO, NLO, NNLO are shown
and compared to a fit involving resummation of $ln(1/x)$ terms ~\cite{thorne}.
One can see that inclusion of such terms 
results in a more reasonable shape for $F_L$.
\begin{figure}{\small Fig.5: Predictions for $F_L$ from conventional DGLAP 
and from low $x$ resummation}
\epsfxsize=15pc
\epsfbox{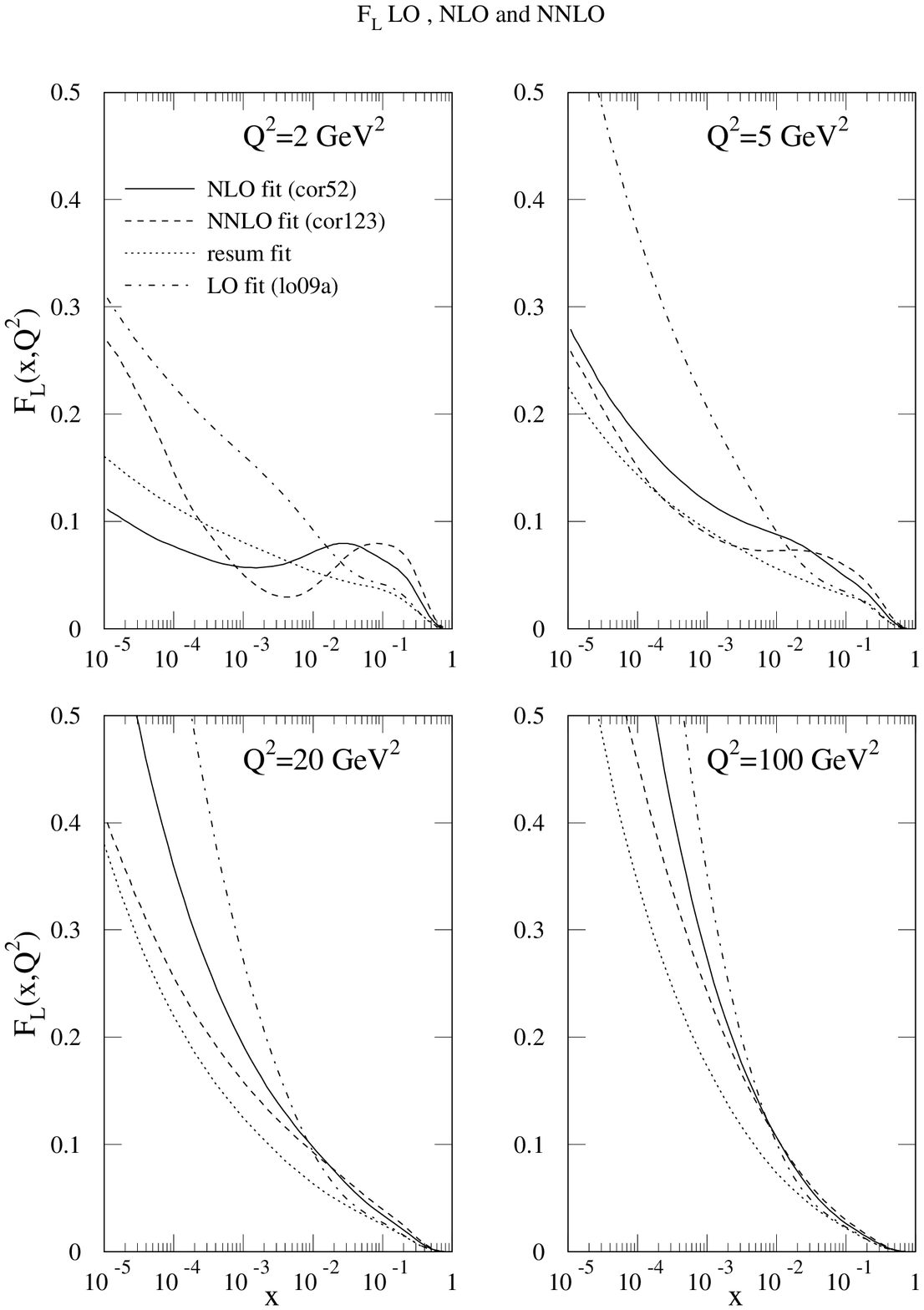}
\label{fig:flrob}
\end{figure}  

Such predictions indicate that measurements of $F_L$ are very important.
A model independent measurement at the interesting low values of $x$ cannot be
done without varying the HERA beam energy~\cite{amcsfl}, but H1 have made a
measurement which depends only on the validity of extrapolation of data on
the reduced cross-section, $\sigma_r = F_2 - y^2/Y^+ F_L$, from
low  $y$, where $F_L$ is not important, to high $y$ (see ref~\cite{h1967} for
details of the method). The measurements, shown in Fig.6, are 
consistent with conventional NLO DGLAP calculations, but presently there
is insufficient precision to discriminate against alternative calculations.
\begin{figure}{\small Fig.6: H1 and fixed target $F_L$ measurements and the 
H1 QCD fit}
\epsfxsize=15pc
\epsfbox{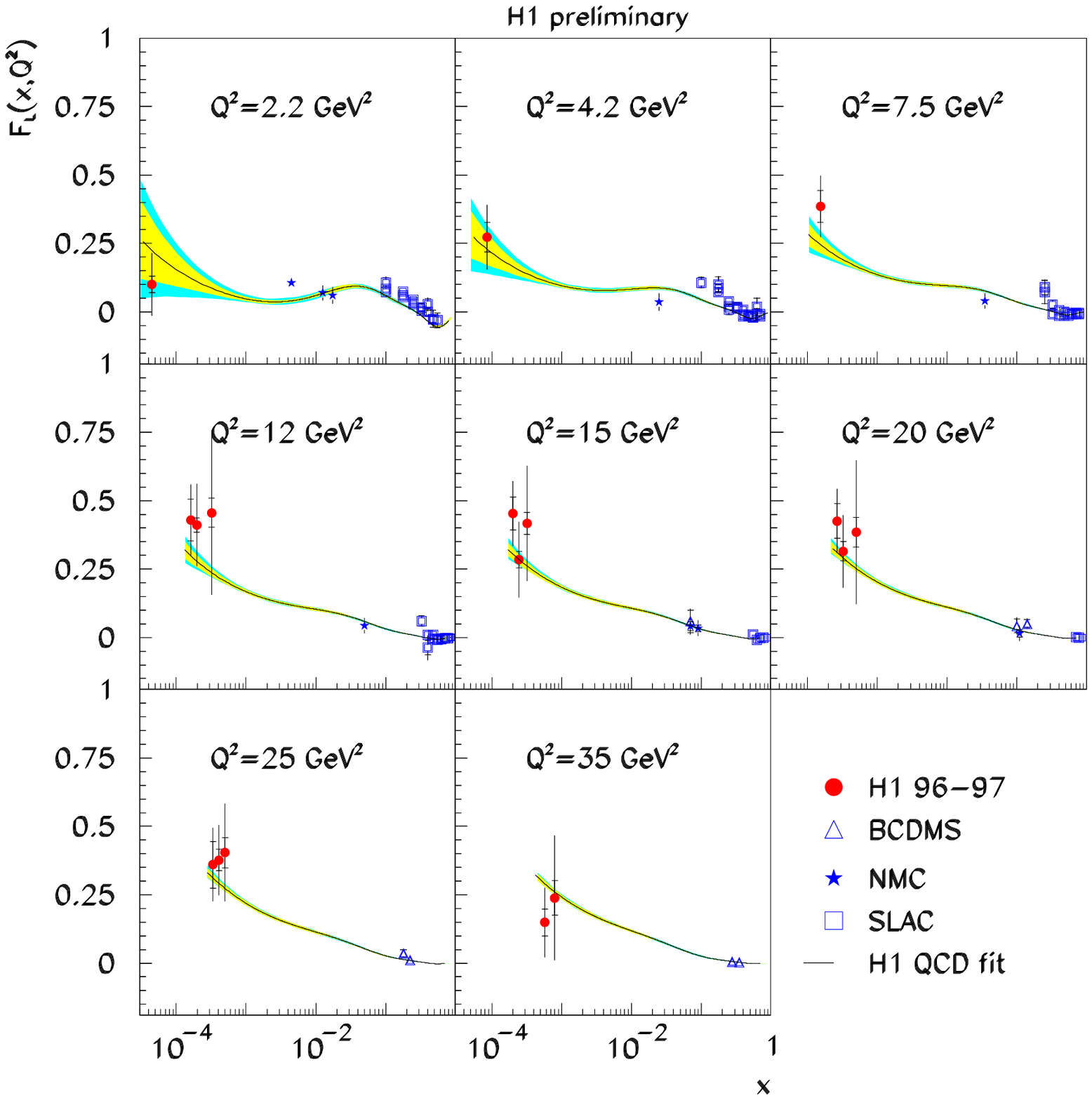}
\label{fig:h1fl}
\end{figure} 

ZEUS has also presented data from their Beam Pipe Tracker (BPT) which enables
measurements in the very low $Q^2$ region~\cite{bpt}.
There has been a lot of work on trying to understand the transition from
non-perturbative physics at $Q^2 \rightarrow 0$ to larger $Q^2$ where pQCD
predictions are valid. Since very low $Q^2$ also means very low $x$, there
are further possible modifications
to conventional theory, when the high parton densities generated at low
$x$ result in the need for non-linear terms in the evolution equations.
Such effects have been termed shadowing and may lead to saturation of the
proton's parton densities~\cite{amcs}. As we have seen, the strong rise of the 
gluon density
at small $x$ is tamed when we go to lower $Q^2$, but the change to a 
valence-like shape may be a feature of our using incorrect evolution equations
in the shadowing regime. Clearly precision data in this regime are very 
important.

In Fig.7 we present the low $Q^2$ data as $F_2$ data as a function of $Q^2$
in fixed $y$ bins. The higher $Q^2$ data are also
shown, so that one can see the shape of the transition. 
\begin{figure}{\small Fig.7: HERA $F_2$ versus $Q^2$ for 
fixed $y$ bins, with QCD and Regge fits}
\epsfxsize=15pc
\epsfbox{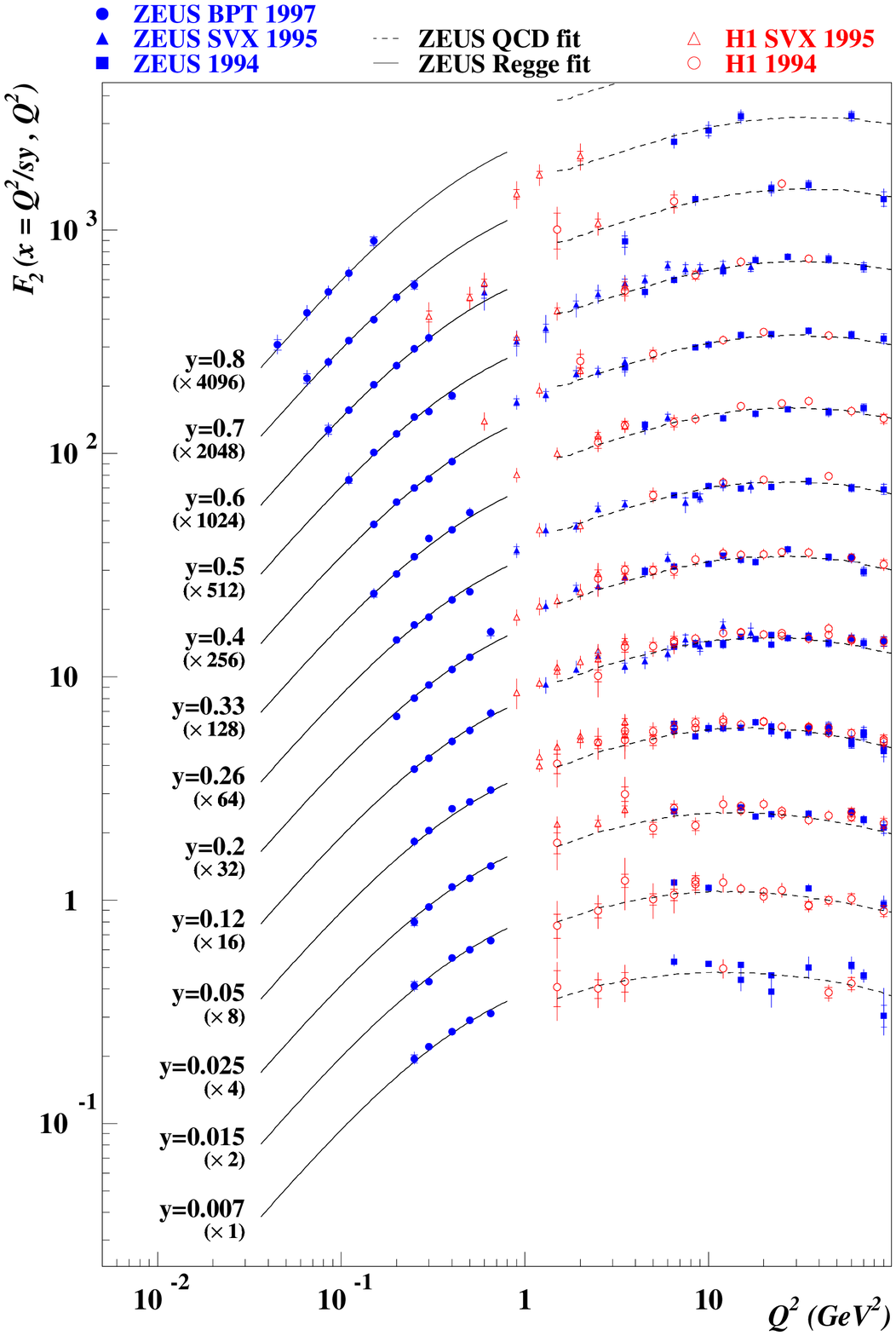}
\label{fig:bpty}
\end{figure} 
At low $x$, the centre of mass energy of the $\gamma^*p$ system is large
($W^2 = Q^2/x$) so that we are in the Regge region for this interaction.
For $Q^2 < 1 GeV^2$, pQCD calculations become
inadequate to describe the shape of the data, so that Regge inspired models 
have been used. These in turn cannot describe data at larger $Q^2$, 
but there have been many 
attempts to extend such models to incorporate QCD effects at higher $Q^2$, see
\cite{amcs}. At low $x$,
\begin{equation}
\sigma^{\gamma^*p}(W^2,Q^2)\approx{4\pi^2\alpha\over Q^2}F_2(x,Q^2)
\end{equation}
relates $F_2$ to the $\gamma^*p$ cross-section. Since we know that the
real photon cross-section at $Q^2 = 0$ is finite, this implies that
 $F_2 \rightarrow 0$ as $Q^2 \rightarrow 0$. Whereas at larger $Q^2$, we know 
that  $F_2$ becomes flattish (baring the QCD 
logarithmic scaling violations). All successful models must predict such a 
transition, but it is now more of a challenge to fit the exact shape of the 
new precision data.

The low $Q^2$ measurements have also been combined with the main data sample to
produce updated plots of $dF_2/dln_{10}Q^2$ versus $x$ and $Q^2$ at fixed $W$,
 see Fig.8.
These plots show a turn over, which moves to lower $Q^2$ and 
higher $x$ as $W$ falls, and this has been interpreted as evidence for dipole 
models of the
transition region which involve parton saturation~\cite{foster}. However, at
low $x$ values,
this derivative is related to the shape of the gluon distribution, and the 
turnover can be fitted by pQCD DGLAP fits, if we believe that the low $Q^2$ 
gluon is really valence-like.
It is also true that if $dF_2/dln_{10}Q^2$ is plotted against $x$ at fixed
$Q^2$ there is no sign of a turnover down to the lowest $Q^2$ values. The
signal of saturation in such a plot would be a change in the slope. Looking
at Fig.9 it is clear
that data of even higher precision would be  necessary to establish this.
\begin{figure}{\small Fig.8: ZEUS $dF_2/dln_{10}Q^2$ data versus $x$ and $Q^2$
 at fixed $W$ values}
\epsfxsize=20pc
\epsfbox{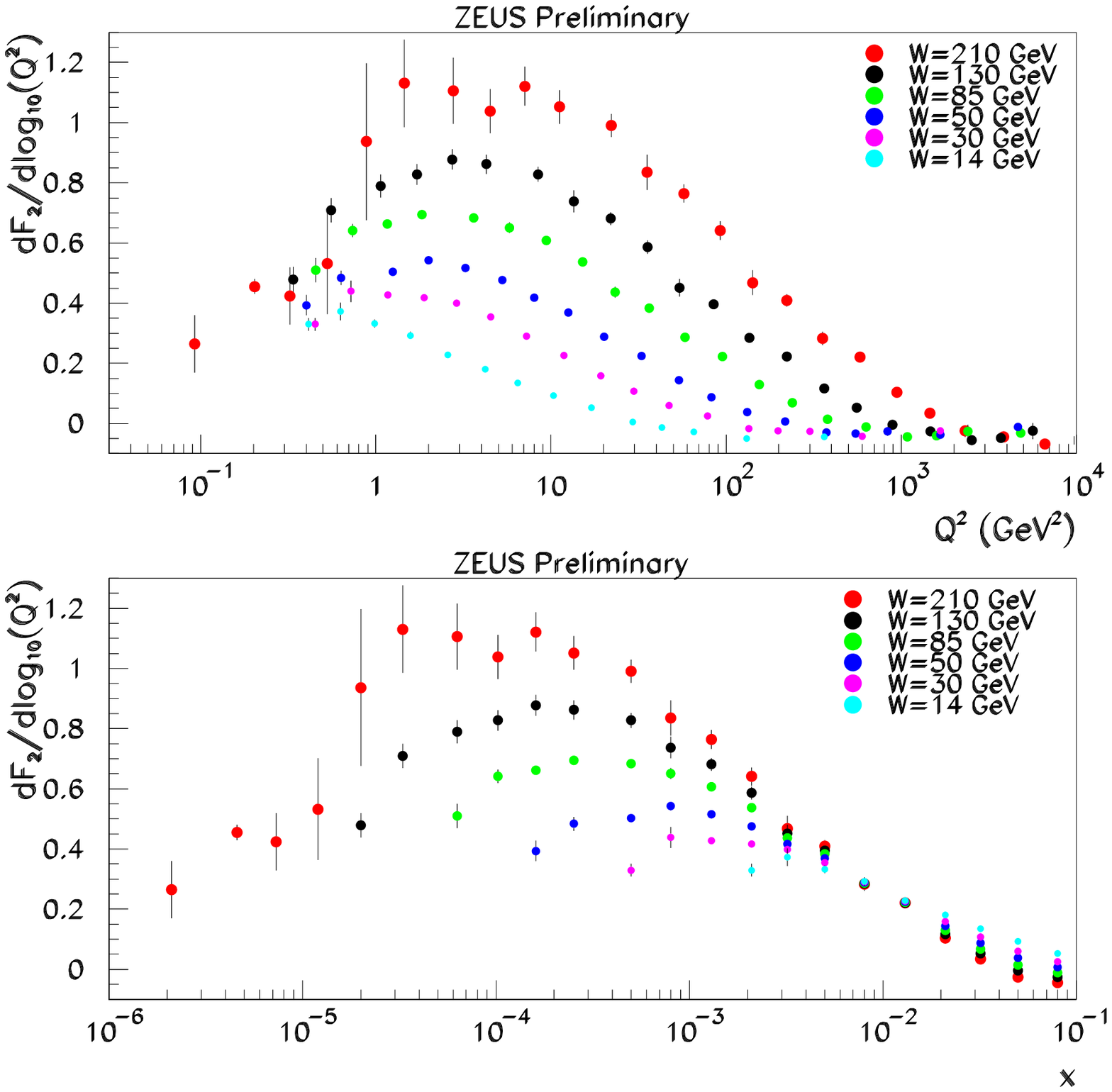}
\label{fig:caldwell}
\end{figure} 
\begin{figure}{\small Fig.9: ZEUS $dF_2/dln_{10}Q^2$ data versus $x$ at fixed $Q^2$
 values}
\epsfxsize=20pc
\epsfbox{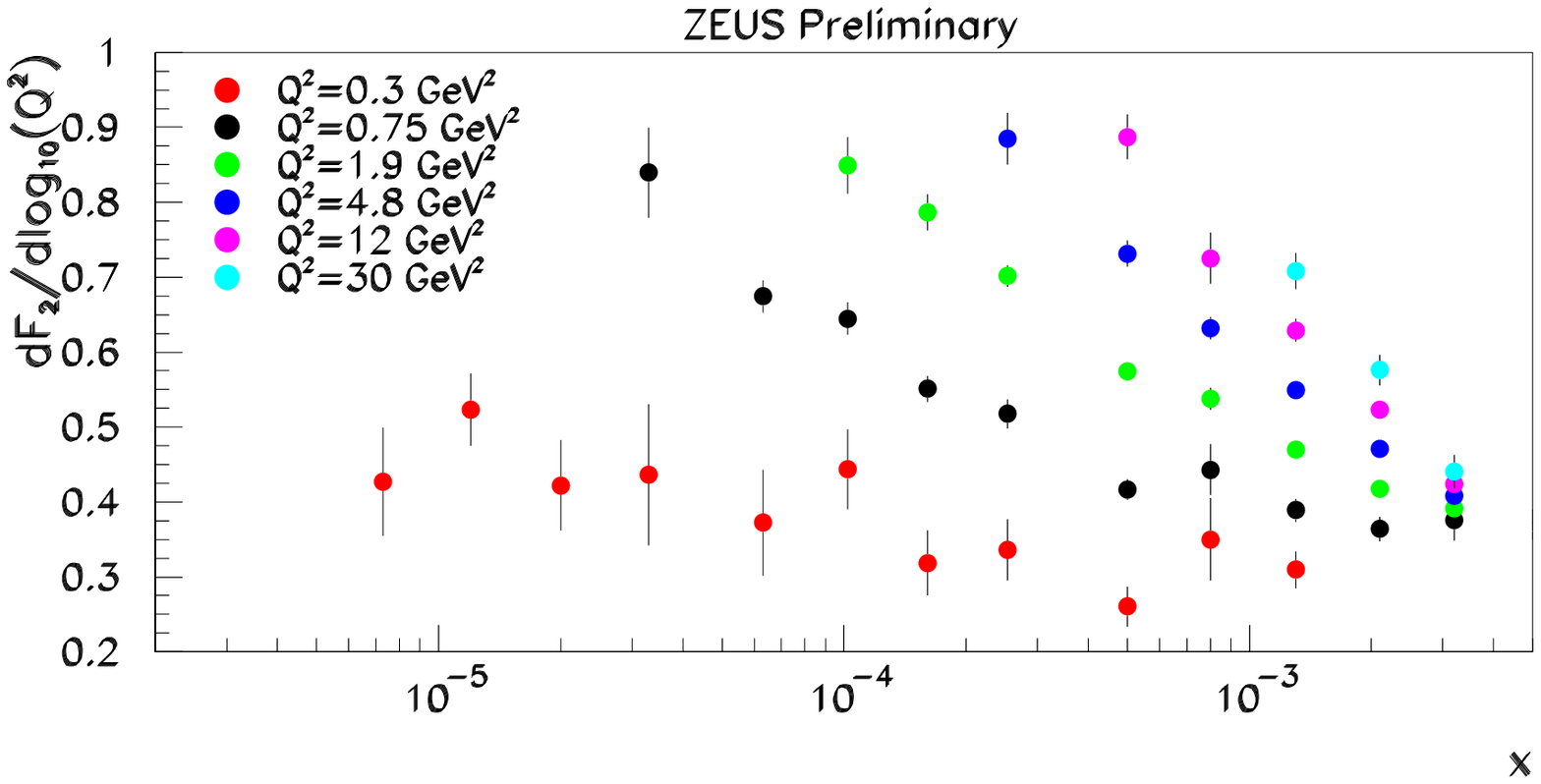}
\label{fig:fixq}
\end{figure}


\end{document}